\documentclass{article}

\usepackage{PRIMEarxiv}

\usepackage[utf8]{inputenc} 
\usepackage[T1]{fontenc}    
\usepackage{hyperref}       
\usepackage{url}            
\usepackage{booktabs}       
\usepackage{amsfonts}       
\usepackage{nicefrac}       
\usepackage{microtype}      
\usepackage{lipsum}
\usepackage{fancyhdr}       
\usepackage{graphicx}       
\graphicspath{{media/}}     

\title{Multiparametric Quantum Sensing of Liquids Using NV Centres and Tethered Magnetic Nanoparticles}

\author{
  Johannes Fiedler, Martin Møller Greve, Justas Zalieckas \\
  Department of Physics and Technology\\
  University of Bergen\\
  All\'egaten 55, 5007 Bergen, Norway\\
  \texttt{johannes.fiedler@uib.no} \\
}

\begin{document}
\maketitle

\begin{abstract}
We propose a new concept for non-invasive, multiparametric liquid analysis based on nitrogen-vacancy (NV) centre magnetometry, relaxometry and surface-tethered magnetic nanoparticles. Magnetic nanoparticles are anchored to a diamond surface via DNA strands, forming nanoscale mechanical oscillators whose thermally driven motion is strongly influenced by the surrounding liquid environment. The resulting time-dependent magnetic fields couple to near-surface NV centres and are detected via optically detected magnetic resonance or changes in spin coherence time. By spatially patterning the diamond surface with regions functionalised by DNA tethers of different lengths, sequences, or chemical modifications, a single liquid is mapped onto a high-dimensional quantum response vector rather than a single scalar observable. Changes in viscosity, molecular adsorption, or chemical interactions modify the dynamics of magnetic nanoparticles in a region-specific manner, enabling differential sensing across the surface. We outline the physical transduction mechanism, discuss relevant scaling relations, and assess experimental feasibility using established wide-field NV magnetometry and relaxometry methods. The proposed platform combines quantum sensing with surface heterogeneity, offering a versatile route toward parallel, label-free liquid characterisation.
\end{abstract}


\section{Introduction}
\label{sec:introduction}

Nitrogen-vacancy (NV) centres in diamond have become a widely used platform for ambient-conditions nanoscale sensing~\cite{Doherty2013,Schirhagl2014}. Their optical accessibility and long spin coherence times enable detection of weak magnetic fields near surfaces and interfaces, making them attractive for applications ranging from condensed-matter systems to biological environments.

In addition to static magnetic fields, near-surface NV centres are sensitive to magnetic noise generated by fluctuating nanoscale systems~\cite{Hall2016,Myers2017}. Such fluctuations can arise from thermal motion, spin dynamics, or other time-dependent processes near the diamond surface. Recent experiments have demonstrated the real-time magnetic detection of DNA-mediated interactions between tethered magnetic nanoparticles and diamond interfaces using wide-field NV-based measurements~\cite{Sun2025}. These developments illustrate that NV-centre measurements can provide access not only to static spatial information, but also to dynamical processes at solid--liquid interfaces.

At the same time, many problems in liquid sensing involve coupled physical and chemical properties that are difficult to characterise with a single observable. Conventional sensing approaches often focus on one parameter, such as viscosity, ion concentration, or the presence of a specific analyte. However, complex liquids and reactive environments frequently exhibit correlated changes across several quantities simultaneously. This naturally motivates sensing concepts that probe a liquid through multiple coupled response channels rather than through a single scalar signal.

In this work, we propose a platform that combines wide-field NV-centre sensing with surface-tethered magnetic nanoparticles for multiparametric liquid sensing. Magnetic nanoparticles attached to the diamond surface via DNA tethers undergo thermally driven motion in the surrounding liquid. The resulting magnetic field fluctuations that are sensitive to the local environment are then detected by nearby NV centres. By deliberately introducing spatially distinct surface regions with different tether and nanoparticle properties, the same liquid can be mapped onto a multidimensional response pattern measured across the surface.

The emphasis of the present work is not on the detailed optimisation of a specific device architecture, but rather on the underlying sensing principle. In the following sections, we introduce the conceptual platform, discuss the physical transduction mechanism and the resulting response space, and outline the experimental feasibility of such a sensing approach.

\section{Conceptual platform}
\label{sec:concept}

\begin{figure*}
  \centering
  \includegraphics[width=0.7\textwidth]{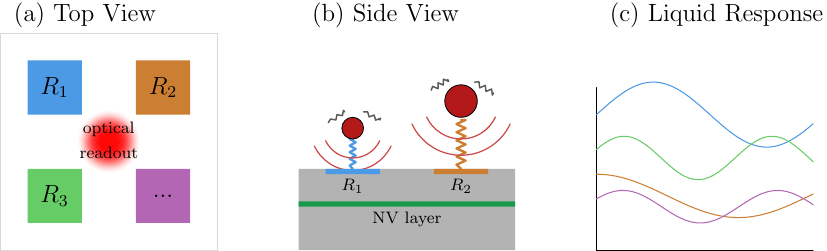}
  \caption{Conceptual platform for multiparametric quantum sensing of liquids. (a) Top view of the diamond sensor surface hosting a near-surface ensemble of nitrogen-vacancy (NV) centres. The surface is patterned into multiple spatial regions (R$_1$, R$_2$, R$_3$, \ldots), each functionalised with distinct DNA tethers that bind magnetic nanoparticles. Wide-field optical excitation and readout enable parallel interrogation of all regions under identical liquid conditions. (b) Side view illustrating representative sensing regions with different tether properties and nanoparticle characteristics. Thermally driven motion of tethered magnetic nanoparticles generates time-dependent magnetic fields that couple to nearby NV centres, modifying their spin resonance and coherence properties. The nanoparticle dynamics depend sensitively on the surrounding liquid environment. (c) Conceptual representation of the resulting sensor response. A single liquid is mapped onto a set of region-specific quantum signals, forming a multidimensional response vector that enables differential and multiparametric liquid characterisation.
}
  \label{fig:concept}
\end{figure*}

The proposed sensing platform, illustrated in Fig.~\ref{fig:concept}, combines wide-field NV-centre sensing with surface-tethered magnetic nanoparticles to enable parallel, multiparametric liquid analysis. The central idea is to transduce liquid-dependent mechanical fluctuations into quantum-detectable magnetic signatures while exploiting surface heterogeneity to generate a multidimensional sensor response. Fig.~\ref{fig:concept}(a) shows a top view of the proposed sensor surface. A near-surface ensemble of NV centres is addressed optically over a wide field of view, enabling simultaneous readout from multiple spatially separated regions. The surface is patterned into distinct sensing areas denoted schematically as R$_1$, R$_2$, R$_3$, \ldots, each functionalised with DNA tethers of different lengths, sequences, chemical modifications, or nanoparticle characteristics. Although all regions are exposed to the same liquid environment, their mechanical response functions differ due to the local surface functionalisation.

The transduction mechanism is illustrated schematically in Fig.~\ref{fig:concept}(b). Magnetic nanoparticles are tethered to the diamond surface via DNA strands and undergo thermally driven motion in the surrounding liquid. These fluctuations generate time-dependent magnetic fields at the position of nearby NV centres, leading to measurable changes in their spin resonance and coherence properties. Since the nanoparticle dynamics depend on the local environment, variations in viscosity, molecular adsorption, hydrodynamic coupling, or chemical interactions modify the measured magnetic noise spectrum.

Related experimental approaches have recently demonstrated the dynamic magnetic detection of DNA hybridisation processes
using NV-centre imaging of tethered magnetic nanoparticles~\cite{Sun2025}. In contrast to monitoring a single interaction channel, the present work focuses on deliberately engineered surface heterogeneity and the resulting multidimensional response space for liquid characterisation.

A central aspect of the platform is the deliberate introduction of surface heterogeneity. Rather than measuring a single observable, the liquid environment is mapped onto a set of region-dependent quantum signals, as illustrated conceptually in Fig.~\ref{fig:concept}(c). Different tether configurations, therefore, respond differently to the same environmental perturbation, producing a multidimensional response pattern across the surface.

In a simplified picture, the sensor response may be represented by a vector
\begin{equation}
    \mathbf{S} = \hat{M}\,\mathbf{L}\,,
\end{equation}
where $\mathbf{L}$ characterises the relevant liquid properties and $\mathbf{S}$ contains the measured responses from the individual sensing regions. The matrix $\hat{M}$ describes how strongly each region responds to a given environmental parameter. In this representation, the effective sensing capacity is determined by the rank of $\hat{M}$. Ideally, $N$ sufficiently distinct sensing regions provide access to an $N$-dimensional response space, enabling several coupled liquid properties to be distinguished simultaneously. In practice, correlations between different regions will reduce the number of independent response channels, making the diversity of the surface functionalisation an important design parameter.

The proposed platform, therefore, does not rely on a single target analyte or sensing modality. Instead, it combines quantum-limited magnetic readout with engineered surface diversity to generate information-rich response patterns that can be analysed comparatively across the sensor array. The concept is compatible with established wide-field NV sensing methods and surface functionalisation techniques, while naturally lending itself to parallel sensing and classification-based approaches.

\section{Physical transduction and scaling considerations}
\label{sec:transduction}

The sensing mechanism proposed here relies on transducing liquid-dependent mechanical fluctuations into a magnetic signature detectable by near-surface NV centres. In this section, we outline the essential physical ingredients and relevant scaling relations, without aiming at a detailed quantitative description of a specific experimental realisation.

The motion of a magnetic nanoparticle tethered to the diamond surface can be described, to leading order, as an overdamped stochastic oscillator. For small displacements around the equilibrium position, the dynamics may be captured by a Langevin equation of the form
\begin{equation}
\gamma \dot{x}(t) + k x(t) = \xi(t)\,,
\end{equation}
where $x(t)$ denotes the particle displacement, $\gamma$ is an effective viscous damping coefficient determined by the surrounding liquid and near-surface hydrodynamics, $k$ is the effective stiffness of the DNA tether, and $\xi(t)$ represents thermal noise. Both $\gamma$ and $k$ are sensitive to the liquid environment, as well as to molecular adsorption, conformational changes of the tether, or modifications of local surface interactions.

The magnetic nanoparticle generates a stray magnetic field that depends on its instantaneous position. For small fluctuations around the equilibrium distance $r_0$ from a nearby NV centre, the magnetic field can be linearised as
\begin{equation}
\delta B(t) \simeq \left.\frac{\partial B}{\partial x}\right|_{r_0} x(t)\,,
\end{equation}
such that the spectral density of magnetic field fluctuations $S_B(\omega)$ directly reflects the mechanical fluctuation spectrum of the tethered nanoparticle. Changes in the liquid environment, therefore, modify the characteristic frequency scale and linewidth of $S_B(\omega)$ through their influence on $\gamma$ and $k$.

Near-surface NV centres are sensitive to such magnetic noise through both their spin relaxation and dephasing dynamics. High-frequency magnetic fluctuations predominantly affect the longitudinal relaxation time $T_1$, while low-frequency and quasi-static fluctuations contribute to dephasing and inhomogeneous broadening characterised by $T_2$ and $T_2^*$. By selecting appropriate measurement protocols, different frequency components of the nanoparticle motion can thus be accessed using the same quantum sensor platform.

Importantly, in the proposed architecture, all surface regions are probed using an identical NV-based readout, while the mechanical response functions differ due to variations in tether properties. As a result, the same liquid-induced modification of $\gamma$ or $k$ leads to distinct changes in the magnetic noise spectra observed across the surface. This parallel mapping of a single environment onto multiple, non-identical response channels forms the basis for the multiparametric sensing concept introduced here.

The simplified treatment presented above is intended to highlight the relevant physical scaling relations rather than to provide a complete quantitative model. More detailed descriptions may incorporate hydrodynamic corrections near solid boundaries, nonlinear tether dynamics, or realistic nanoparticle magnetic field profiles. However, the essential transduction pathway from liquid properties to quantum-detectable magnetic signature remains unchanged.

The relevant dynamical timescales of the tethered nanoparticles are expected to lie in a frequency range accessible to established NV-based relaxometry and dephasing measurements. Depending on tether stiffness, nanoparticle size, and liquid viscosity, characteristic fluctuation frequencies are expected to range from the kHz to MHz regime, overlapping with the sensitivity window of near-surface NV centres. Importantly, the proposed sensing concept does not rely on exceptionally long coherence times, but rather on differential changes in the magnetic noise spectrum measured across multiple sensing regions.

A detailed discussion of NV-based magnetic noise spectroscopy and near-surface sensing can be found in Refs.~\cite{Doherty2013,Schirhagl2014}.

Although the platform is compatible with conventional NV-based magnetometry, the dominant sensing mechanism considered here is based on fluctuation-induced changes in the local magnetic noise environment. The relevant observables are therefore primarily the NV relaxation and dephasing dynamics, rather than static magnetic-field shifts.

\section{Experimental feasibility and outlook}
\label{sec:feasibility}

The proposed sensing platform relies exclusively on well-established experimental techniques, each individually well established and feasible within current NV-based quantum sensing architectures. Wide-field optical excitation and readout of near-surface NV ensembles has been demonstrated extensively, providing parallel access to spatially resolved magnetic signals over areas ranging from micrometres to millimetres. Similarly, magnetic nanoparticles with well-controlled size and magnetic moment are routinely employed in nanoscale sensing and actuation experiments, while surface functionalisation using DNA tethers is a mature technique in both biophysical and nanotechnological contexts.

From an implementation perspective, the platform requires the guided immobilisation of magnetic nanoparticles on a diamond surface via molecular tethers, as well as the confinement of a liquid layer above the surface. Both requirements are compatible with existing surface-chemistry protocols, DNA-spotting techniques, and microfluidic approaches. Recent experiments have demonstrated the stable wide-field detection of DNA-tethered magnetic nanoparticles at diamond interfaces under physiological conditions~\cite{Sun2025}, supporting the experimental plausibility of the sensing architecture proposed here. The use of wide-field optical readout avoids the need to individually address surface regions, enabling parallel interrogation under identical environmental conditions. Importantly, the concept does not rely on a specific nanoparticle size, tether length, or liquid composition, but instead exploits relative and differential responses across multiple regions, relaxing constraints on absolute calibration.

The introduction of surface heterogeneity plays a central role in the robustness of the sensing approach. By design, different regions respond differently to the same environmental perturbation, allowing changes in liquid properties to be identified by characteristic patterns rather than by absolute signal levels. This differential strategy is expected to mitigate the effects of global drifts, laser-intensity fluctuations, and slow changes in NV properties that affect all regions similarly.

Looking ahead, the proposed platform naturally lends itself to extensions beyond static liquid characterisation. Time-dependent measurements could enable real-time monitoring of evolving liquid environments, adsorption processes, and surface chemical reactions. Increasing the number and diversity of functionalised regions would further enrich the accessible response space, opening the door to classification-based approaches and data-driven analysis strategies. More detailed theoretical modelling and targeted experimental studies may refine the quantitative interpretation of the observed signals, but are not prerequisites for demonstrating the core sensing concept.

Overall, the combination of quantum-limited magnetic readout, engineered surface heterogeneity, and parallel interrogation offers a versatile route toward multiparametric liquid sensing. The conceptual framework outlined here provides a foundation for future experimental implementations and highlights how quantum sensors can be leveraged not only for high sensitivity but also for rich, information-dense characterisation of complex environments.

\bibliographystyle{unsrt}  
\bibliography{references}

\end{document}